\documentclass[12pt,a4paper]{article}

\usepackage{ifthen}
\newboolean{pdflatex}
\setboolean{pdflatex}{true}

\newboolean{articletitles}
\setboolean{articletitles}{true}

\newboolean{uprightparticles}
\setboolean{uprightparticles}{false}

\newboolean{inbibliography}
\setboolean{inbibliography}{false}

\usepackage[top=1in, bottom=1.25in, left=1in, right=1in]{geometry}

\usepackage{microtype}
\usepackage{lineno}
\usepackage{xspace}
\usepackage{caption}

\usepackage{graphicx}
\usepackage{color}
\usepackage{colortbl}

\usepackage{amsmath}
\usepackage{amssymb}
\usepackage{amsfonts}
\usepackage{upgreek}

\newcommand*\patchAmsMathEnvironmentForLineno[1]{
\expandafter\let\csname old#1\expandafter\endcsname\csname #1\endcsname
\expandafter\let\csname oldend#1\expandafter\endcsname\csname
end#1\endcsname
 \renewenvironment{#1}
   {\linenomath\csname old#1\endcsname}
   {\csname oldend#1\endcsname\endlinenomath}
}
\newcommand*\patchBothAmsMathEnvironmentsForLineno[1]{
  \patchAmsMathEnvironmentForLineno{#1}
  \patchAmsMathEnvironmentForLineno{#1*}
}
\AtBeginDocument{
\patchBothAmsMathEnvironmentsForLineno{equation}
\patchBothAmsMathEnvironmentsForLineno{align}
\patchBothAmsMathEnvironmentsForLineno{flalign}
\patchBothAmsMathEnvironmentsForLineno{alignat}
\patchBothAmsMathEnvironmentsForLineno{gather}
\patchBothAmsMathEnvironmentsForLineno{multline}
\patchBothAmsMathEnvironmentsForLineno{eqnarray}
}

\usepackage{hyperref}
\usepackage[all]{hypcap}

\usepackage{xspace}
\usepackage{upgreek}

\def\lhcb {\mbox{LHCb}\xspace}

\def\MagUp {\mbox{\em Mag\kern -0.05em Up}\xspace}

\ifthenelse{\boolean{uprightparticles}}
{

 \def\Peta        {\ensuremath{\upeta}\xspace}

 \def\Pmu         {\ensuremath{\upmu}\xspace}

 \def\Ppsi        {\ensuremath{\uppsi}\xspace}

 \def\PDelta      {\ensuremath{\Delta}\xspace}
 \def\PXi      {\ensuremath{\Xi}\xspace}
 \def\PLambda      {\ensuremath{\Lambda}\xspace}
 \def\PSigma      {\ensuremath{\Sigma}\xspace}
 \def\POmega      {\ensuremath{\Omega}\xspace}
 \def\PUpsilon      {\ensuremath{\Upsilon}\xspace}

 \def\PB      {\ensuremath{\mathrm{B}}\xspace}
 
 \def\PD      {\ensuremath{\mathrm{D}}\xspace}

 \def\PJ      {\ensuremath{\mathrm{J}}\xspace}
 \def\PK      {\ensuremath{\mathrm{K}}\xspace}

 \def\Pb      {\ensuremath{\mathrm{b}}\xspace}
 \def\Pc      {\ensuremath{\mathrm{c}}\xspace}

 \def\Pi      {\ensuremath{\mathrm{i}}\xspace}

 \def\Ps      {\ensuremath{\mathrm{s}}\xspace}

}
{

 \def\Peta        {\ensuremath{\eta}\xspace}

 \def\Pmu         {\ensuremath{\mu}\xspace}

 \def\Ppsi        {\ensuremath{\psi}\xspace}
 
 \mathchardef\PDelta="7101
 \mathchardef\PXi="7104
 \mathchardef\PLambda="7103
 \mathchardef\PSigma="7106
 \mathchardef\POmega="710A
 \mathchardef\PUpsilon="7107
 
 \def\PB      {\ensuremath{B}\xspace}
 
 \def\PD      {\ensuremath{D}\xspace}

 \def\PJ      {\ensuremath{J}\xspace}
 \def\PK      {\ensuremath{K}\xspace}

 \def\Pb      {\ensuremath{b}\xspace}
 \def\Pc      {\ensuremath{c}\xspace}

 \def\Pi      {\ensuremath{i}\xspace}

 \def\Ps      {\ensuremath{s}\xspace}

}

\makeatletter
\ifcase \@ptsize \relax
  \newcommand{\miniscule}{\@setfontsize\miniscule{4}{5}}
\or
  \newcommand{\miniscule}{\@setfontsize\miniscule{5}{6}}
\or
  \newcommand{\miniscule}{\@setfontsize\miniscule{5}{6}}
\fi
\makeatother

\DeclareRobustCommand{\optbar}[1]{\shortstack{{\miniscule (\rule[.5ex]{1.25em}{.18mm})}
  \\ [-.7ex] $#1$}}

\def\mup        {{\ensuremath{\Pmu^+}}\xspace}
\def\mun        {{\ensuremath{\Pmu^-}}\xspace}

\def\squark    {{\ensuremath{\Ps}}\xspace}

\def\cquark    {{\ensuremath{\Pc}}\xspace}

\def\bquark    {{\ensuremath{\Pb}}\xspace}

  \def\Kbar    {{\kern 0.2em\overline{\kern -0.2em \PK}{}}\xspace}

\def\KorKbar    {\kern 0.18em\optbar{\kern -0.18em K}{}\xspace}

  \def\Dbar    {{\kern 0.2em\overline{\kern -0.2em \PD}{}}\xspace}
\def\D       {{\ensuremath{\PD}}\xspace}

\def\DorDbar    {\kern 0.18em\optbar{\kern -0.18em D}{}\xspace}

\def\Dsp     {{\ensuremath{\D^+_\squark}}\xspace}
\def\Dsm     {{\ensuremath{\D^-_\squark}}\xspace}

\def\B       {{\ensuremath{\PB}}\xspace}
\def\Bbar    {{\ensuremath{\kern 0.18em\overline{\kern -0.18em \PB}{}}}\xspace}

\def\BorBbar    {\kern 0.18em\optbar{\kern -0.18em B}{}\xspace}
\def\Bz      {{\ensuremath{\B^0}}\xspace}

\def\Bu      {{\ensuremath{\B^+}}\xspace}

\def\Bp      {{\ensuremath{\Bu}}\xspace}

\def\Bd      {{\ensuremath{\B^0}}\xspace}
\def\Bs      {{\ensuremath{\B^0_\squark}}\xspace}
\def\Bsb     {{\ensuremath{\Bbar{}^0_\squark}}\xspace}

\def\jpsi     {{\ensuremath{{\PJ\mskip -3mu/\mskip -2mu\Ppsi\mskip 2mu}}}\xspace}

  \def\Y#1S{\ensuremath{\PUpsilon{(#1S)}}\xspace}

\def\Lbar        {{\ensuremath{\kern 0.1em\overline{\kern -0.1em\PLambda}}}\xspace}
\def\LorLbar    {\kern 0.18em\optbar{\kern -0.18em \PLambda}{}\xspace}

\def\to                 {\ensuremath{\rightarrow}\xspace}

\def\CP                {{\ensuremath{C\!P}}\xspace}

\newcommand{\phis}{{\ensuremath{\phi_{\squark}}}\xspace}

\def\AT#1     {\ensuremath{A_{\mathrm{T}}^{#1}}\xspace}

\def\C#1      {\ensuremath{\mathcal{C}_{#1}}\xspace}
\def\Cp#1     {\ensuremath{\mathcal{C}_{#1}^{'}}\xspace}
\def\Ceff#1   {\ensuremath{\mathcal{C}_{#1}^{\mathrm{(eff)}}}\xspace}
\def\Cpeff#1  {\ensuremath{\mathcal{C}_{#1}^{'\mathrm{(eff)}}}\xspace}
\def\Ope#1    {\ensuremath{\mathcal{O}_{#1}}\xspace}
\def\Opep#1   {\ensuremath{\mathcal{O}_{#1}^{'}}\xspace}

\newcommand{\tev}{\ifthenelse{\boolean{inbibliography}}{\ensuremath{~T\kern -0.05em eV}\xspace}{\ensuremath{\mathrm{\,Te\kern -0.1em V}}}\xspace}
\newcommand{\gev}{\ensuremath{\mathrm{\,Ge\kern -0.1em V}}\xspace}
\newcommand{\mev}{\ensuremath{\mathrm{\,Me\kern -0.1em V}}\xspace}
\newcommand{\kev}{\ensuremath{\mathrm{\,ke\kern -0.1em V}}\xspace}
\newcommand{\ev}{\ensuremath{\mathrm{\,e\kern -0.1em V}}\xspace}
\newcommand{\gevc}{\ensuremath{{\mathrm{\,Ge\kern -0.1em V\!/}c}}\xspace}
\newcommand{\mevc}{\ensuremath{{\mathrm{\,Me\kern -0.1em V\!/}c}}\xspace}
\newcommand{\gevcc}{\ensuremath{{\mathrm{\,Ge\kern -0.1em V\!/}c^2}}\xspace}
\newcommand{\gevgevcccc}{\ensuremath{{\mathrm{\,Ge\kern -0.1em V^2\!/}c^4}}\xspace}
\newcommand{\mevcc}{\ensuremath{{\mathrm{\,Me\kern -0.1em V\!/}c^2}}\xspace}

\def\mum  {\ensuremath{{\,\upmu\mathrm{m}}}\xspace}

\def\invfb   {\ensuremath{\mbox{\,fb}^{-1}}\xspace}

\def\ps   {\ensuremath{{\mathrm{ \,ps}}}\xspace}
\def\fs   {\ensuremath{\mathrm{ \,fs}}\xspace}

\newcommand{\stat}{\ensuremath{\mathrm{\,(stat)}}\xspace}
\newcommand{\syst}{\ensuremath{\mathrm{\,(syst)}}\xspace}

\newcommand{\chisq}{\ensuremath{\chi^2}\xspace}

\newcommand{\chisqip}{\ensuremath{\chi^2_{\text{IP}}}\xspace}

\def\gsim{{~\raise.15em\hbox{$>$}\kern-.85em
          \lower.35em\hbox{$\sim$}~}\xspace}
\def\lsim{{~\raise.15em\hbox{$<$}\kern-.85em
          \lower.35em\hbox{$\sim$}~}\xspace}

\def\ptot       {\mbox{$p$}\xspace}
\def\pt         {\mbox{$p_{\mathrm{ T}}$}\xspace}

\def\evtgen     {\mbox{\textsc{EvtGen}}\xspace}

\def\geant      {\mbox{\textsc{Geant4}}\xspace}

\def\photos     {\mbox{\textsc{Photos}}\xspace}

\def\pythia     {\mbox{\textsc{Pythia}}\xspace}

\def\tell1  {TELL1\xspace}
\def\ukl1   {UKL1\xspace}

\usepackage{cite}
\usepackage{mciteplus}
\usepackage{longtable}

\begin{document}

\renewcommand{\thefootnote}{\fnsymbol{footnote}}
\setcounter{footnote}{1}

\begin{titlepage}
\pagenumbering{roman}

\vspace*{-1.5cm}
\centerline{\large EUROPEAN ORGANIZATION FOR NUCLEAR RESEARCH (CERN)}
\vspace*{1.5cm}
\noindent
\begin{tabular*}{\linewidth}{lc@{\extracolsep{\fill}}r@{\extracolsep{0pt}}}
\vspace*{-2.7cm}\mbox{\!\!\!\includegraphics[width=.14\textwidth]{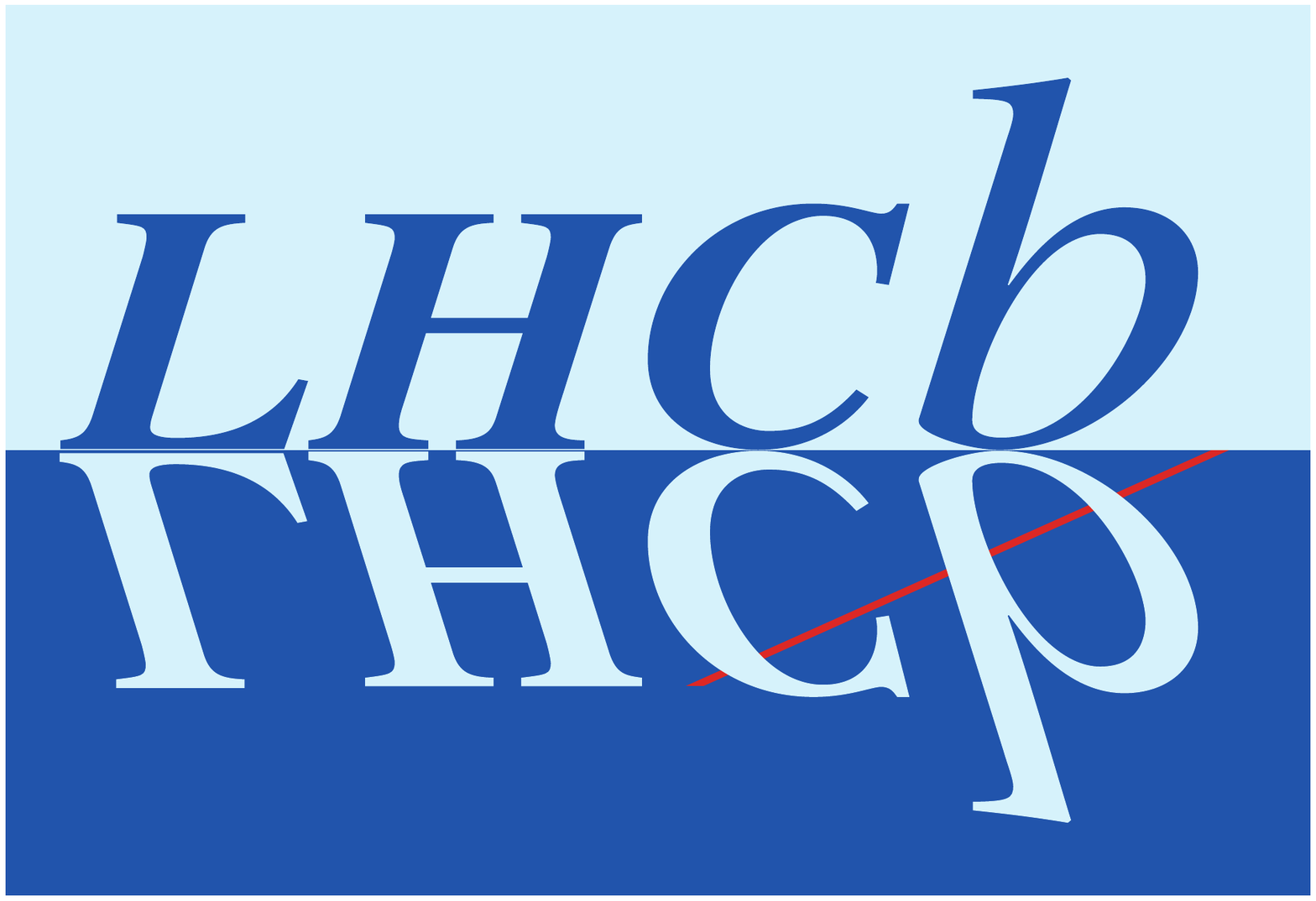}} & &
\\
 & & CERN-EP-2016-166 \\
 & & LHCb-PAPER-2016-017 \\
 & & \today \\
\end{tabular*}

\vspace*{4.0cm}

{\normalfont\bfseries\boldmath\huge
\begin{center}
  Measurement of the $\Bs \rightarrow \jpsi \eta$ lifetime
\end{center}
}

\vspace*{1.0cm}

\begin{center}
The LHCb collaboration\footnote{Authors are listed at the end of this
  paper.} \\
\vspace{0.5cm}
This paper is dedicated to the memory of our friend and colleague
Ailsa Sparkes.
\end{center}
\vspace{0.5cm}
\vspace{\fill}

\begin{abstract}
  \noindent
 Using a data set corresponding to an integrated luminosity of $3 \invfb$,
collected by the LHCb experiment in $pp$ collisions at
centre-of-mass energies of 7 and 8 TeV, the effective lifetime in the
$\Bs \rightarrow \jpsi \eta$ decay mode, $\tau_{\textrm{eff}}$, is measured to be
\begin{displaymath}
\tau_{\textrm{eff}} = 1.479 \pm 0.034~\textrm{(stat)} \pm 0.011 ~\textrm{(syst)} \ps. \nonumber
\end{displaymath}
Assuming \CP conservation, $\tau_{\textrm{eff}}$ corresponds to the
lifetime of the light $\Bs$ mass eigenstate.  This is the first
measurement of the effective lifetime in this decay mode.
\end{abstract}

\vspace*{2.0cm}

\begin{center}
Accepted by Phys. Lett. B.
\end{center}

\vspace{\fill}

{\footnotesize
\centerline{\copyright~CERN on behalf of the \lhcb collaboration, licence \href{http://creativecommons.org/licenses/by/4.0/}{CC-BY-4.0}.}}
\vspace*{2mm}

\end{titlepage}

\newpage
\setcounter{page}{2}
\mbox{~}

\cleardoublepage

\renewcommand{\thefootnote}{\arabic{footnote}}
\setcounter{footnote}{0}

\pagestyle{plain}
\setcounter{page}{1}
\pagenumbering{arabic}


\section{Introduction}
\label{sec:Introduction}
Studies of $B^0_s - \overline{B}^0_s$ mixing provide important
tests of the Standard Model (SM) of particle physics.
In the SM, mixing occurs via box diagrams. Extensions to the SM may introduce
additional \CP-violating phases that alter the value of
the $\Bs - \Bsb$ mixing weak phase, $\phi_s$, from that of the
SM \cite{LHCb-PAPER-2012-031}.
The $\Bs$ system exhibits a sizeable difference in  the decay widths $\Gamma_{\rm L}$
and  $\Gamma_{\rm H}$, where $\rm L$ and $\rm H$ refer to the light and heavy $\Bs$  mass eigenstates, respectively.
The effective lifetime, $\tau_{\textrm{eff}}$, of a $\Bs$ meson decay mode is measured by approximating the decay time
distribution, determined in the $\Bs$ rest system, by a single exponential function. For final states that can be accessed by both
$\Bs$ and $\Bsb$ mesons the effective lifetime  depends on their \CP
components  and is also sensitive to $\phi_s$ \cite{Fleischer:2011cw,
  Fleischer:2011ib}.

In this analysis  $\tau_{\textrm{eff}}$ is determined for the \CP-even $B^0_s \rightarrow
\jpsi \eta$ decay mode. As $\phis$ is measured to be small \cite{LHCb-PAPER-2014-059,LHCb-PAPER-2014-019}
the mass eigenstates are also \CP eigenstates to better than per mille
level and $\tau_{\textrm{eff}}$  measured in $B^0_s \rightarrow
\jpsi \eta$ decays is equal, to good approximation, to the
lifetime of the light $\Bs$ mass eigenstate,  $\tau_{\rm L} \propto \Gamma_{\rm L}^{-1}$. In the
SM $\tau_{\rm L}$ is predicted to be $1.43 \pm 0.03 \ps$~\cite{Lenz:2012mb}.
Measurements of $\tau_{\rm L}$ have previously been reported by
LHCb in the $\Bs \rightarrow \Dsp \Dsm$ and $\Bs \rightarrow K^+ K^-$ decay modes \cite{LHCb-PAPER-2013-060,LHCb-PAPER-2014-011}.
The latter  is dominated by
penguin diagrams, which could arise within and beyond the SM
and gives rise to direct \CP violation.
This then leads to a  different $\tau_{\textrm{eff}}$,
when compared to measurements in
the  $\Bs \rightarrow \Dsp \Dsm$ and $B^0_s \rightarrow
\jpsi \eta$ decays which are mediated by tree diagrams.
Improved  precision on the effective lifetimes $\tau_{\rm L}$ and
$\tau_{\rm H}$ will enable more stringent tests of the consistency
between direct measurements of the decay width difference
$\Delta\Gamma_s = \Gamma_{\rm L} - \Gamma_{\rm H}$  measured in $\Bs \rightarrow \jpsi \phi$ decays and those inferred using effective lifetimes.

The measurement of the effective $\Bs \rightarrow \jpsi \eta$ lifetime presented in this Letter
uses 3~$\invfb$ of data collected in $pp$ collisions at
centre-of-mass energies of 7~TeV and
8~TeV during 2011 and 2012 using the LHCb detector. The $\jpsi$ meson
is reconstructed via the dimuon decay mode and the $\eta$ meson via the diphoton
decay mode. The presence of only two charged particles in the final
state minimizes systematic uncertainties related to the tracking system.

\section{Detector and simulation}
\label{sec:Detector}
The \lhcb detector~\cite{Alves:2008zz,LHCb-DP-2014-002} is a single-arm forward
spectrometer covering the \mbox{pseudorapidity} range $2<\eta <5$,
designed for the study of particles containing \bquark or \cquark
quarks. The detector includes a high-precision tracking system
consisting of a silicon-strip vertex detector surrounding the $pp$
interaction region, a large-area silicon-strip detector (TT) located
upstream of a dipole magnet with a bending power of about
$4{\mathrm{\,Tm}}$, and three stations of silicon-strip detectors and straw
drift tubes placed downstream of the magnet.
The tracking system provides a measurement of momentum, \ptot, of charged particles with
a relative uncertainty that varies from 0.5\,\% at low momentum to 1.0\,\%
at 200\gevc.  Large samples of  $\jpsi\to\mup\mun$
and $B^+ \to \jpsi K^+$ decays, collected
concurrently with the data set used here, were used to calibrate the
momentum scale of the spectrometer to a precision of $0.03\,\%$
\cite{LHCb-PAPER-2012-048}. The minimum distance of a track to a primary vertex (PV), the impact parameter (IP),
is measured with a resolution of $(15+29/\pt)\mum$,
where \pt is the component of the momentum transverse to the beam, in\,\gevc.

Different types of charged hadrons are distinguished using information
from two ring-imaging Cherenkov detectors.
Photons, electrons and hadrons are identified by a calorimeter system consisting of
scintillating-pad and preshower detectors, an electromagnetic
calorimeter and a hadronic calorimeter. The calorimeter response is calibrated
using samples of $\pi^0 \rightarrow \gamma \gamma$ decays. For this
analysis a further calibration was made using  the decay $\eta \rightarrow \gamma
\gamma$,  which results in a precision of $0.07\,\%$ on the neutral energy scale. Muons are identified by a
system composed of alternating layers of iron and multiwire
proportional chambers.

The online event selection is performed by a
trigger~\cite{LHCb-DP-2012-004},
which consists of a hardware stage, based on information from the calorimeter and muon
systems, followed by a software stage, where  a full event
reconstruction is made. Candidate events are required to pass the hardware trigger,
which selects muon and dimuon candidates with high $\pt$ based upon
muon system information. The subsequent
software trigger is composed of two stages. The first performs a
partial event reconstruction and requires events to have two
well-identified oppositely charged muons with an invariant mass larger
than $2.7 \gevcc$. The second stage performs a
full event reconstruction.  Events are retained for further
processing if they contain a displaced $\jpsi \rightarrow \mu^+ \mu^-$
candidate. The decay vertex is required to be well separated
from each reconstructed PV of the proton-proton
interaction by requiring the distance between the PV and the $\jpsi$
decay vertex  divided by its
uncertainty to be greater than three. This introduces a non-uniform
efficiency for $\bquark$-hadron candidates that have a decay time less than
$0.1 \ps$.

Simulated $pp$ collisions are generated using
\pythia~\cite{Sjostrand:2006za,*Sjostrand:2007gs}  with a specific
\lhcb configuration~\cite{LHCb-PROC-2010-056}.  Decays of hadronic
particles are described by \evtgen~\cite{Lange:2001uf}, in which
final-state radiation is generated using
\photos~\cite{Golonka:2005pn}. The interaction of the generated
particles with the detector, and its response, are implemented using the \geant toolkit~\cite{Allison:2006ve,
  *Agostinelli:2002hh} as described in Ref.~\cite{LHCb-PROC-2011-006}.

\section{Selection}
\label{sec:selection}
A two-step procedure, is used to optimize the selection of $\Bs \rightarrow
\jpsi \eta$ decay candidates. These studies use simulated data samples together with the
high mass sideband of the data ($5650 < m(\jpsi \eta) < 5850 \mevcc$),
which is not used in the subsequent determination of $\tau_{\textrm{eff}}$. In
a first step, loose selection criteria are applied that
reduce background significantly whilst retaining
high signal efficiency. Subsequently, a
multivariate selection (MVA) is used to reduce further the combinatorial
background. This is optimized using pseudoexperiments to obtain the best precision on the measured
$\Bs$ lifetime.

The selection starts from a pair of oppositely charged particles,
identified as muons, that form a common decay vertex. Combinatorial
background is suppressed by requiring that \chisqip
of the muon candidates\footnote{The quantity \chisqip is defined as the difference between
  the \chisq of the PV reconstructed with and without the considered
  particle.} to all reconstructed PVs to be larger than four. To ensure a high reconstruction
efficiency the muon candidates are required to have a pseudorapidity between
2.0 and 4.5. The invariant mass of the dimuon candidate must be within $50 \mevcc$ of the known $\jpsi$ mass
\cite{PDG2014}. In addition, the trigger requirement
that the $\jpsi$ decay length divided by its uncertainty is greater
than three is reapplied.

Photons are selected from neutral clusters reconstructed in the electromagnetic
calorimeter~\cite{LHCb-DP-2014-002} that have a transverse
energy in excess of $300~\mathrm{MeV}$ and a confidence level to
be a photon, $\mathcal{P}_{\gamma}$,  greater than 0.009. The latter requirement has an
efficiency of $98\,\%$ for the simulated signal sample whilst removing $23\,\%$
of the background in the high mass sideband. To suppress combinatorial
background, if either of the photons when combined with any other
photon candidate in the event has an invariant mass within
$25~\mathrm{MeV}/c^2$ of the known $\pi^0$ meson mass~\cite{PDG2014}
the candidate is rejected.

Candidate $\eta\to\gamma \gamma$ decays are selected from
diphoton combinations with an invariant mass within
$70~\mathrm{MeV}/c^2$ of the known $\eta$~mass \cite{PDG2014} and with a transverse
momentum larger than $2 \gevc$. The
decay angle between the photon momentum in the $\eta$~rest frame and the
direction  of Lorentz boost from the laboratory frame to
the $\eta$~rest frame, $\theta^*_{\Peta}$, is required to satisfy $\left| \cos
  \theta^*_{\Peta} \right| < 0.8$.

The  $\jpsi$ and $\eta$ candidates are combined to form candidate $\B^0_{(s)}$ mesons.
The average number of PVs in each event is 1.8 (2.0) for the 2011 (2012) dataset.
When multiple PVs are reconstructed, the one with the minimum \chisqip
to the $B^0_{(\squark)}$ candidate is chosen. A kinematic fit is performed to improve the invariant mass resolution~\cite{Hulsbergen:2005pu}.
In this fit the momentum vector of the $\B^0_{(s)}$ candidate is constrained to point to the PV
and the intermediate resonance masses are constrained to their known values.
The reduced $\chi^2$~of this fit, $\chi^2/\mathrm{ndf}$, is required to be less than five.
The measured $\B^0_{(\squark)}$ decay time must be larger than $0.3 \ps$ and less than $10 \ps$.
If more than one PV is reconstructed in an event the properties of the unassociated vertices are studied.
Any candidate for which there is a second PV with $\chisqip < 50$ is
rejected. This requirement has an efficiency of $98 \%$ that is almost
flat as a function of the decay time and reduces background due to incorrect association of the $\B^0_{(\squark)}$ candidate to a PV.
Finally, as in Ref.~\cite{LHCb-PAPER-2013-065}, the position of the PV along the beam-line is required to be within
$10$~cm of the nominal interaction point, where the standard deviation of this variable is approximately $5 \,$cm.
This criterion leads to a 10~\% reduction in signal yield but defines a fiducial region where the
reconstruction efficiency is uniform.

The second step of the selection process is based on a neural network \cite{Hocker:2007ht}, which is trained
using the simulated signal sample and the high-mass sideband of the data for background.
Seven variables that show good agreement between data and simulation
and that do not significantly bias the $\B^0_{(\squark)}$
decay time distribution are used to train the neural net:
the $\chi^2$/{\rm ndf} of the kinematic fit; the $\pt$ of the $B^0_{(\squark)}$ and $\eta$ mesons;
the minimum $\pt$ of the two photons; $\left| \cos \theta^*_{\Peta} \right|$; the minimum $\mathcal{P}_{\gamma}$ of
the two photons and the total hit multiplicity in the TT sub-detector.

The requirement on the MVA output was chosen to minimize the
statistical uncertainty on the fitted $\tau_{\textrm{eff}}$
using a sample of 100 pseudoexperiments. The chosen value removes $94 \, \%$ of background candidates whilst retaining $69 \, \%$
of the signal candidates.
After applying these requirements $2 \, \%$ of events contain multiple candidates from which only one,
chosen at random, is kept.
\enlargethispage{\baselineskip}

\section{Fit model}
\label{sec:fitmodel}
The effective lifetime is determined by performing a two-dimensional maximum likelihood fit to the
unbinned distributions of the $\B^0_{(s)}$ candidate invariant mass
and decay time
\begin{displaymath}
t = \frac{m \cdot l}{p},
\end{displaymath}
where $l$ is the candidate decay length, $p$ the candidate momentum
and $m$ the reconstructed invariant mass of the candidate. The fit is performed for candidates with $ 5050 <m(\jpsi
\eta) < 5650 \mevcc$ and $0.3 < t < 10 \ps $.
The fit model has four components: the $\Bs \rightarrow \jpsi \eta$ signal, background from the
$\Bz \rightarrow \jpsi \eta$ decay, background from partially reconstructed
$\Bs \rightarrow \jpsi\eta X$ decays, and combinatorial background.

In the fit, the decay-time distribution of each component is convolved with a Gaussian resolution
function whose width is fixed to the standard deviation of the
decay-time resolution in simulated data. A decay-time acceptance
function accounts for the dependence of the signal efficiency on
several effects. The procedure used to model the decay-time acceptance
is described in detail in Ref. \cite{LHCb-PAPER-2013-065}.
The overall acceptance, $A_{\textrm{tot}}$, is factorised into the product of the selection ($A_{\textrm{sel}}$), trigger
($A_{\textrm{trig}}$) and vertex ($A_{\beta}$) acceptance functions, determined as described below.
The effect of the selection requirements, dominated by the cut on the
displacement of the muons from the PV, is studied using simulation and
parameterised with the form
\begin{equation*}
A_{\textrm{sel}} = \frac{1 - c_0  t} {1 + (c_1 t)^{-c_2}},
\end{equation*}
where $t$ is the decay time, and $c_{0}, c_{1}$ and $c_{2}$ are parameters
determined from the simulation and summarized in Table \ref{tab:selacc}.
\begin{table}[h]
\caption{\small Acceptance parameters due to the selection
  requirements ($A_{\textrm{sel}}$). The correlation coefficients are
  $\rho_{c_0 c_1} = 0.51 $,  $\rho_{c_0 c_2} = 0.62 $ and  $\rho_{c_1 c_2} = 0.95$.}
\begin{center}
\renewcommand{\arraystretch}{1.2}
\small
\begin{tabular}{c|c}
Parameter                          &    Value  \\
\hline
      $c_0$               &    $(6.5 \pm 0.4) \times 10^{-3} \ps^{-1}$ \\
      $c_1$                      &    $(6.6 \pm 0.3)
                                                       \ps^{-1}$ \\
      $c_2$                      &    $1.50\pm 0.04$ \\
\end{tabular}
\label{tab:selacc}
\end{center}
\end{table}
In the second level of the software trigger a cut is applied on the decay length significance of the $\jpsi$
candidate, which biases the decay time distribution. The trigger efficiency,
$A_{\textrm{trig}}$, is measured separately for the 2011 and 2012
datasets using events that are selected by a dedicated prescaled trigger in which this requirement is
removed. It increases approximately linearly from $98 \%$ at $t = 0.3
\ps$ to $100 \%$ $4 \ps$. The resulting acceptance shape is parameterised in bins of
decay time with linear interpolation between the bins. Finally, the
reconstruction efficiency of the vertex detector decreases
as the distance of closest approach of the decay products to the
$pp$ beam-line increases. This effect is studied using $\Bp \rightarrow
\jpsi K^+$ decays where the kaon is reconstructed without using
vertex detector information \cite{LHCb-PAPER-2013-065} and parameterised with
the form
\begin{equation*}
A_{\beta} = 1 - \beta  t - \gamma  t^2,
\end{equation*}
where the parameters $\beta$ and $\gamma$ are determined separately
for the 2011 and 2012 data. The obtained values are summarized in
Table \ref{tab:gamma}.
\begin{table}[h]
\caption{\small Values of the $\beta$ and $\gamma$ factor fitting the
  quadratic form. The first uncertainty is
statistical and the second from the propagation of the uncertainty on
the efficiency versus the distance of closest approach obtained with
the $\Bp \rightarrow \jpsi K^+$ calibration sample. The correlation
coefficienct between $\beta$ and $\gamma$ is 0.8. }
\begin{center}
\small
\renewcommand{\arraystretch}{1.5}
\begin{tabular}{l|c|c}
Sample & $\beta$  [\%]  & $\gamma$ [\%] \\ \hline
2011 data  & $ 0.39 \pm 0.06 ^{-0.01}_{+0.07}$ &  $0.115 \pm
0.021^{-0.004}_{+0.001}$ \\
2012 & $ 0.93 \pm 0.080 ^{+0.001}_{-0.01}  $ & $0.051 \pm 0.023 ^{-0.006}_{+0.006}$ \\
\end{tabular}
\label{tab:gamma}
\end{center}
\end{table}

\begin{figure}[b!]
\begin{center}
\resizebox{5.0in}{!}{\includegraphics{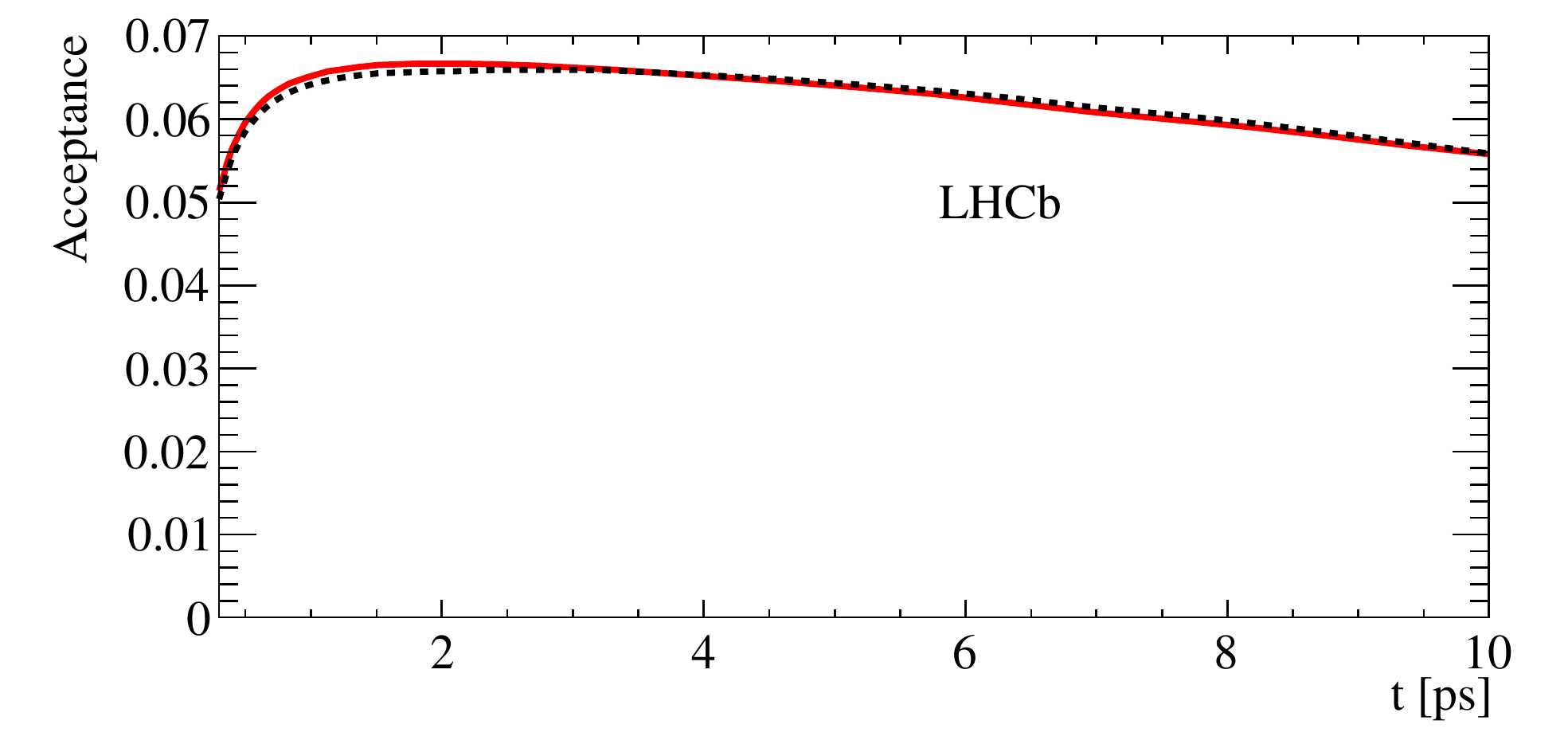}}
\caption{\small Total acceptance function, $A_{\textrm{tot}}$ for 2011 data (black dashed
  line) and 2012 data (solid red). }
\label{fig:accdata}
\end{center}
\end{figure}

Figure~\ref{fig:accdata} shows the overall acceptance curves obtained for the 2011
and 2012 datasets. The shape of $A_{\textrm{tot}}$ is mainly determined
by  $A_{\textrm{sel}}$,  whose uncertainty is dominated by the
size of the simulation sample. The overall acceptance correction is
relatively small. Fitting the simulated data with and without the
correction $\tau_{\textrm{eff}}$ changes by $13\fs$.

The invariant mass distribution for the $\Bs \rightarrow \jpsi \eta$
signal is parameterised by a Student's t-distribution. The
Bukin~\cite{Bukin} and JohnsonSU~\cite{JOHNSON01061949} functions are
considered for systematic variations. In the fit to the data, the
shape parameters of this
distribution are fixed to the simulation values. The decay time distribution for this component is modelled with an exponential function
convolved with the detector resolution and multiplied by the detector
acceptance, as discussed above.

The second component in the fit accounts for the $\Bz \rightarrow \jpsi \eta$ decay.
As the invariant mass resolution is approximately $48 \mevcc$ this overlaps with the $\Bs$ signal mode.
Its mass distribution is modelled, analogously to the $\Bs$ component, with a Student's t-distribution,
with resolution parameters fixed to values determined in the simulation.
The mass difference between the $\Bs$ and $\Bz$ mesons, and the $\Bz$
lifetime, are fixed to their known central values:
$m(\Bs) - m(\Bd) =  87.29 \pm 0.26 \mevcc$~\cite{LHCb-PAPER-2015-010} and
$\tau(\Bz) = 1.519 \pm 0.005 \ps$~\cite{PDG2014} and the uncertainty
propagated to the systematic error. Similarly, the relative yield of the $\Bz$ and $\Bs$ components, $f_r$, is fixed to $(7.3 \pm 0.8) \, \%$ calculated from
the average of the branching fractions measurements made by the Belle~\cite{Chang:2006sd,Chang:2012gnb} and
LHCb collaborations~\cite{LHCb-PAPER-2014-056}, and the measured fragmentation
fractions~\cite{LHCb-PAPER-2011-018,LHCb-PAPER-2012-037,LHCb-CONF-2013-011}.

Combinatorial background is modelled by a first order Chebyshev polynomial in mass and
the sum of two exponentials in decay time. In the fit to the data the
lifetime of the shorter lived component is fixed to the value found in
the fit to the sideband. As a systematic
variation of the mass model, an exponential function is considered.

Background from partially reconstructed decays of $\bquark$ hadrons is
studied using a simulated $\bquark \overline{\bquark}$ sample. Using
this sample an additional background component, due to partially reconstructed
$\Bs \rightarrow \jpsi\eta X$ decays, is identified. Background from
this source lies at invariant masses below $5100 \mevcc$ and has a lifetime of $1.33 \pm 0.10 \ps$. This component is
modelled by a Novosibirsk function \cite{Ikeda:1999aq} in mass and
an exponential in time. All parameters for this component apart from
the yield are fixed to the simulation values in the fit to the data.

The fit has eight free parameters: the yield of the $\Bs
\rightarrow \jpsi \eta$ component ($N^{\Bs}$), the combinatorial
background yield ($N^{\textrm{comb}}$), the partially reconstructed background
yield ($N^{\textrm{partial}}$), the $\Bs$ mass, the lifetime of the signal
component ($\tau_{\textrm{\textrm{eff}}}$), the coefficient of the combinatorial background component in
mass ($a_{\textrm{comb}}$), the longer lived background lifetime ($\tau_{\textrm{comb}}$) and the
fraction of the short-lived background ($f_{\textrm{comb}}$). Independent fits
are performed for the 2011 and 2012 data and a weighted average of
the two lifetime values is made. The correctness of the fit procedure
is validated using the full simulation and pseudoexperiments. No
significant bias is found and the uncertainties estimated by the fit
are found to be accurate.

\section{Results}
\label{sec:results}
\begin{figure}[b!]
\begin{center}
\includegraphics[width=0.49\textwidth]{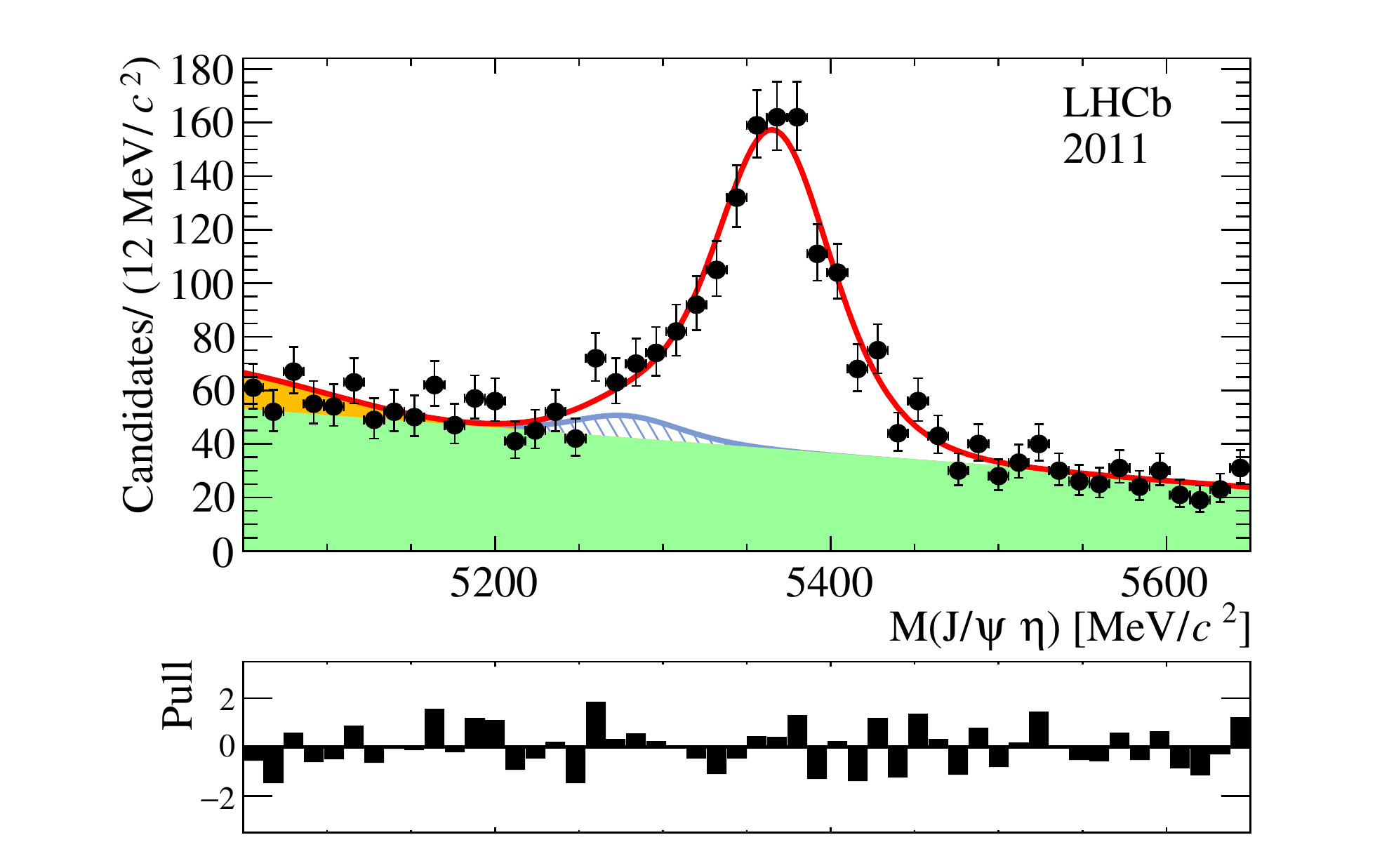}
\includegraphics[width=0.49\textwidth]{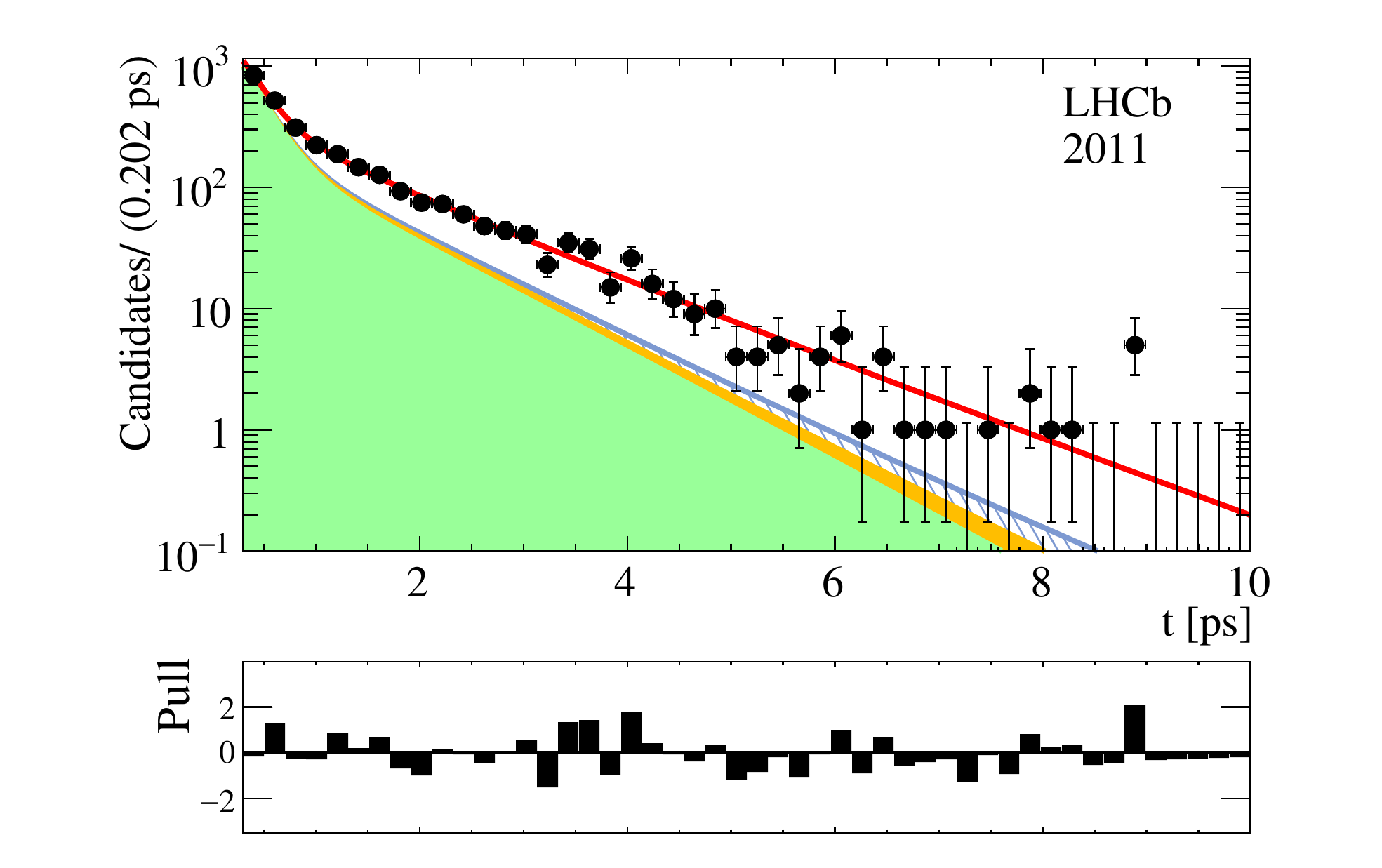}
\includegraphics[width=0.49\textwidth]{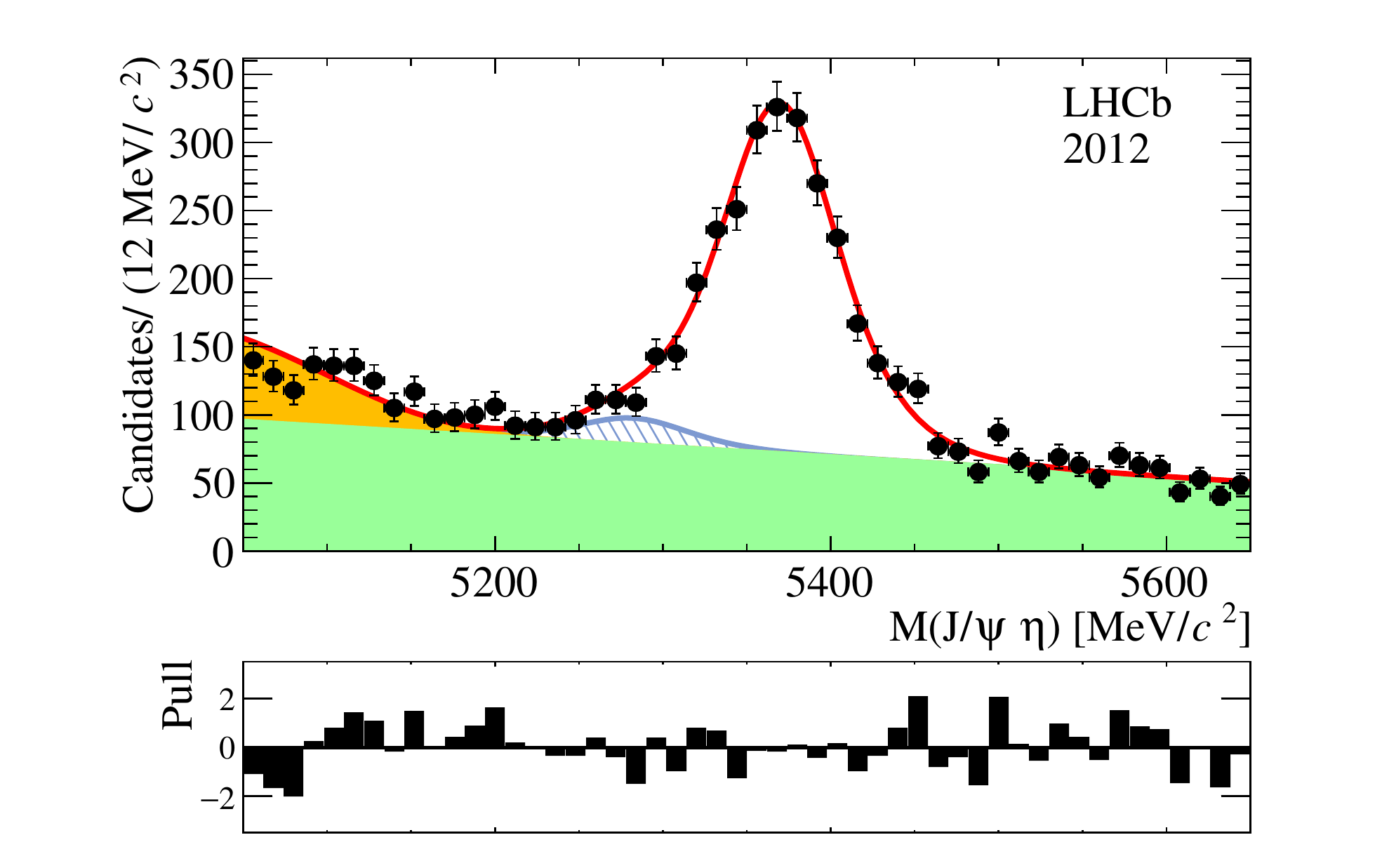}
\includegraphics[width=0.49\textwidth]{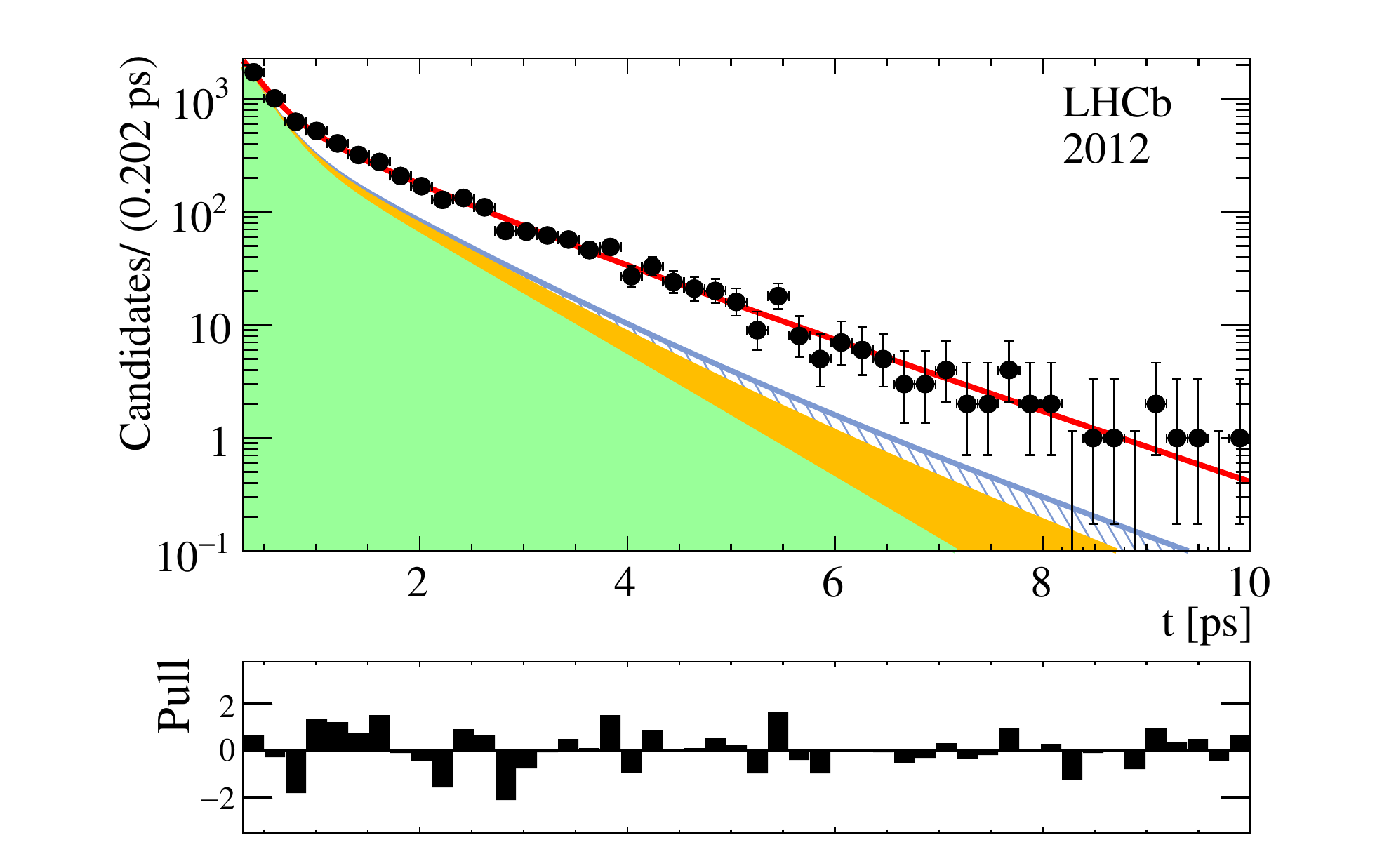}
\caption{\small Mass and decay time distributions for
  the 2011 dataset (top row) and 2012 dataset (bottom row). The fit model
  described in the text is superimposed (red line). The
  partially reconstructed component is shown in solid yellow (dark grey),
  the combinatorial background in solid green (light grey)
  and the $\Bz$ component as open blue. The pull, \textit{i.e.} the difference between the observed and fitted value divided by the uncertainty,
  is shown below each of the plots. }
\label{fig:fits}
\end{center}
\end{figure}

Figure \ref{fig:fits} shows the
fit projections in mass and decay time for the 2011 and 2012 data. The
corresponding fit results are summarized in Table
\ref{tab:massFit}. The fitted signal yields of the two years scale
according to the known integrated luminosity and $\bquark$-hadron
production cross-section. There is some tension in the relative
yield of the partially reconstructed background between the two
years. However, this parameter is almost uncorrelated with
$\tau_{\textrm{eff}}$ and this tension has no impact on the result.
The average of the fitted values of $\tau_{\textrm{eff}}$ is
\begin{displaymath}
\tau_{\textrm{eff}} = 1.479 \pm 0.034 \ps \nonumber,
\end{displaymath}
where the uncertainty is statistical.

\renewcommand{\arraystretch}{1.3}
\begin{table}[h]
\caption{\small Parameters of the fit to $B^0_{(s)} \rightarrow \jpsi \eta$ candidates for the
  2011 and 2012 datasets. Uncertainties are statistical only.}
\begin{center}
\small
\begin{tabular}{l|r@{$\,\pm\,\,$}l|r@{$\,\pm\,\,$} l}
                                              &
                                              \multicolumn{4}{c}{Fitted value} \\
\raisebox{1.5ex}[-1.5ex]{Fit parameter}                          &
\multicolumn{2}{c}{2011}  &  \multicolumn{2}{c}{2012}   \\ \hline
$N^{\Bs}$                       &    960 & 42 & 2061 & 60  \\
$m_{\Bs}$ [\mevcc]             &  5365.6  & 1.8  & 5369.6 & 1.3  \\
$\tau_{\textrm{eff}}$ [ps]         &  1.485 & 0.060  & 1.476 & 0.041 \\
$N^{\textrm{comb}} $                           &          1898 & 64   &  3643 & 89      \\
$N^{\textrm{partial}}$                            &         81 & 26  & 345 & 39            \\
$a_{\textrm{comb}}$                              & $-0.37$ & 0.05     &   $-0.31$ & 0.03        \\
$f_{\textrm{comb}}$  &  0.52 & 0.03 & 0.49 & 0.02 \\
$\tau_{\textrm{comb}}$ [ps] &  0.97 & $0.06$ & 0.82 & $0.04$\\
\end{tabular}
\label{tab:massFit}
\end{center}
\end{table}

The main source of systematic uncertainty is due to the
modelling of the decay time acceptance function (Section
\ref{sec:fitmodel}). Varying the parameters of the acceptance function within their correlated
uncertainties, a variation of the fitted lifetime of 10 fs is found, which is
assigned as a systematic uncertainty. Uncertainties on $A_{\textrm{sel}}$ due
to the parameterisation of this effect are evaluated to be
negligible by replacing the functional form with a histogram.
The statistical and systematic uncertainties on $A_{\beta}$ are
evaluated by repeating the fit and varying the parameterisation within
its uncertainties.  The statistical uncertainty on $A_{\textrm{trig}}$ is propagated by
generating an ensemble of histograms with each bin varied within
its statistical uncertainty. Systematic uncertainties on $A_{\textrm{trig}}$ are estimated to be small by varying the
binning of the histogram and considering an alternative analytic
form. In simulation studies the efficiency of the MVA is found to be
independent of the decay time within uncertainties. Conservatively,
allowing for a linear dependence, an uncertainty of $1.7 \fs$ is assigned.

The influence of the decay time resolution is estimated by increasing
its value from 51 to $70 \fs$. This variation covers the variation of
the resolution with decay time and any possible discrepancy in the
resolution between data and simulation. The change in
$\tau_{\textrm{eff}}$ from this variation is negligible. The impact of
the uncertainties in $f_r$, the $\Bs-\Bz$ mass splitting, and
the $\Bz$ lifetime are evaluated by repeating the fit procedure
varying these parameters within their quoted uncertainties.

Further uncertainties arise from the modelling of the time
distributions of the background components. In the default fit the
lifetime of the short-lived component is fixed to the value found in a fit
to the mass sideband. Removing this constraint changes the
result by $4 \fs$, which is assigned as a systematic uncertainty. The uncertainty due to the fixed lifetime of
the partially reconstructed component is found to be negligible.

Uncertainties arising from the modelling of the signal and
background mass distributions are evaluated using the discrete
profiling method described in Ref.~\cite{Dauncey:2014xga} and found
to be negligible. Further small uncertainties arise due to the limited
knowledge of the length scale of the detector along the beam axis ($z-$scale), the charged particle momentum scale and
the neutral particle energy scale.

The stability of the result has been tested against a number of
possible variations, such as the fitted invariant mass range, the requirement on the
IP of the muons, the MVA requirement and analysing the sample according to the number of reconstructed PVs.
No significant change in the final result is found and hence no further systematic uncertainty is assigned.

All the uncertainties are summarized in
Table~\ref{tab:systematics}. Adding them in quadrature leads to a
total systematic uncertainty of $11.1 \fs$ which is dominated by the size of
the simulation sample used to determine the acceptance and to validate
the analysis procedure.
\renewcommand{\arraystretch}{1.1}
\begin{table}[t]
\caption{\small Systematic uncertainties on the lifetime
  measurement. Uncertainties less than $0.1 \fs$ are indicated by a dash.}
\label{tab:systematics}
\begin{center}
\small
\begin{tabular}{l|c}
Source &  Uncertainty [$\fs$] \\
\hline
$A_{\textrm{sel}}$ &  10.0 \\
$A_\beta$ (stat)  & 2.0 \\
$A_\beta$ (syst)  & 0.1 \\
$A_{\textrm{trig}}$ (stat) & 0.6 \\
$A_{\textrm{trig}}$ (syst) & 0.6 \\
MVA &  1.7 \\
Time resolution &  -- \\
$f_r$  & 1.2 \\
$\Bs - \Bz$ mass difference &  -- \\
$\Bz$ lifetime & 0.2 \\
Releasing $\tau_{\textrm{back}}$ & 4.0 \\
Varying $\tau_{\textrm{partial}}$ &  -- \\
Mass model & -- \\
Momentum scale                 &  --  \\
$z$-scale         & 0.3 \\ \hline
Total & 11.1 \\
\end{tabular}
\end{center}
\end{table}

\section{Summary}
\label{sec:summary}
\begin{figure}[h]
\begin{center}
\resizebox{5in}{!}{\includegraphics{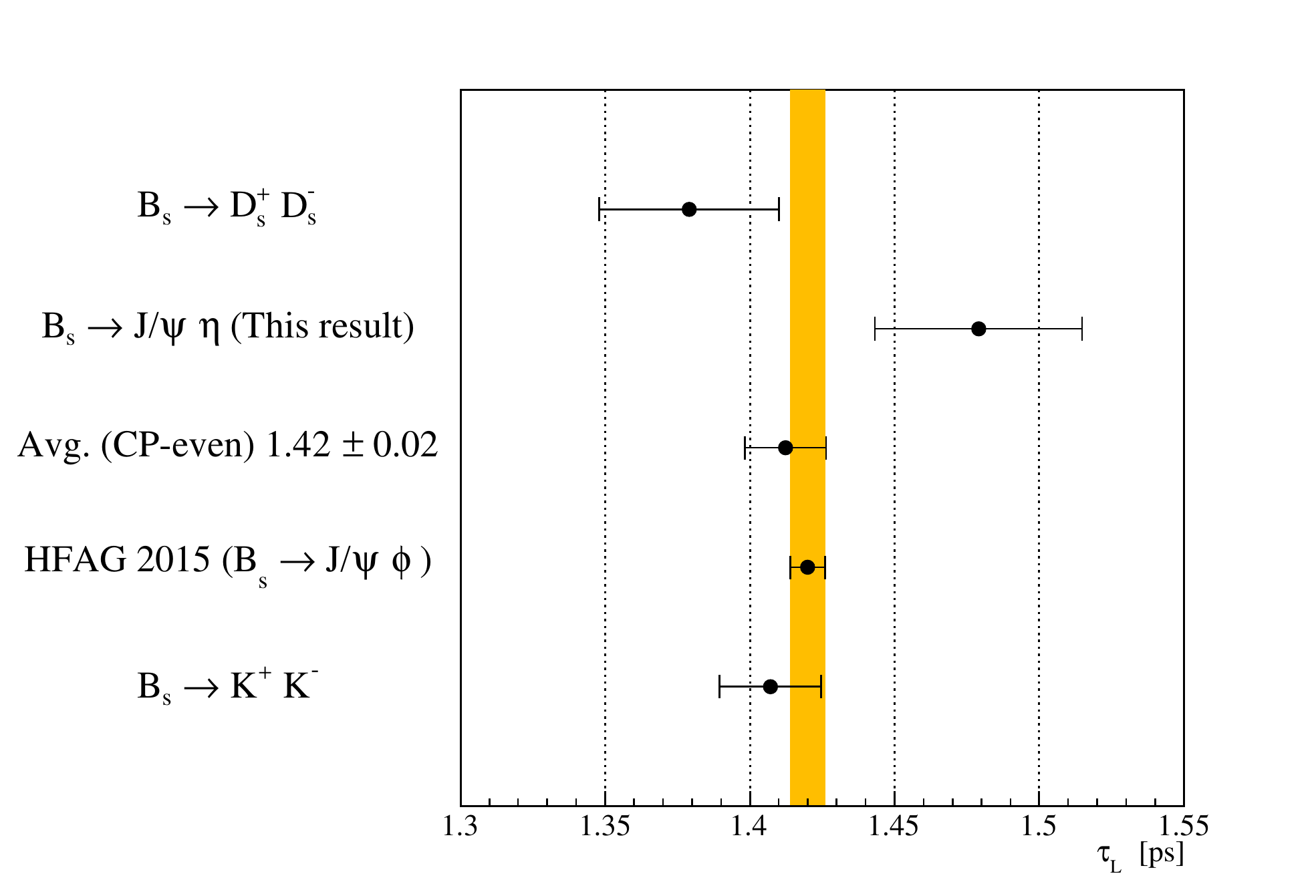}}
\caption{\small Summary of measurements of $\tau_{\rm L}$. The yellow band
  corresponds to the 2015 HFAG central value and uncertainty. }
\label{fig:comp}
\end{center}
\end{figure}

Using data collected by LHCb, the effective lifetime in the $B^0_s
\rightarrow \jpsi \eta$ decay
mode is measured to be
\begin{displaymath}
\tau_{\textrm{eff}} = 1.479 \pm 0.034~\textrm{(stat)} \pm 0.011 ~\textrm{(syst)} \ps. \nonumber
\end{displaymath}
In the limit of $\CP$ conservation, $\tau_{\textrm{eff}}$ is equal to
the lifetime of the light $\Bs$ mass eigenstate $\tau_{L}$.
The present measurement is consistent with, and has similar precision to, the effective lifetime determined
using the $\Bs \rightarrow \Dsp \Dsm$ decay mode~\cite{LHCb-PAPER-2013-060},
$\tau_{\textrm{eff}}(\Dsp \Dsm) = 1.379 \pm 0.026 \stat \pm 0.017 \syst$\ps and also with the value measured
in the $\Bs \rightarrow K^+ K^-$ mode~\cite{LHCb-PAPER-2014-011},
$\tau_{\textrm{eff}}(K^+ K^-) = 1.407 \pm 0.016 \stat \pm 0.007 \syst$\ps, where penguin diagrams are expected to
be more important.
Averaging the tree level measurements gives $\tau_{\rm eff} = 1.42 \pm 0.02 \ps$ in good agreement with the
expectations of the Standard Model~\cite{Lenz:2012mb}, $\tau_{\rm L} = 1.43 \pm 0.03 \ps$ and the value quoted
by HFAG~\cite{HFAG} from measurements made in the $B^0_s \rightarrow \jpsi\phi$ mode, $\tau_{\rm L} = 1.420 \pm 0.006 \ps$.
The values from these different measurements are compared in Fig.~\ref{fig:comp}.

\section*{Acknowledgements}
\noindent We express our gratitude to our colleagues in the CERN
accelerator departments for the excellent performance of the LHC. We
thank the technical and administrative staff at the LHCb
institutes. We acknowledge support from CERN and from the national
agencies: CAPES, CNPq, FAPERJ and FINEP (Brazil); NSFC (China);
CNRS/IN2P3 (France); BMBF, DFG and MPG (Germany); INFN (Italy);
FOM and NWO (The Netherlands); MNiSW and NCN (Poland); MEN/IFA (Romania);
MinES and FANO (Russia); MinECo (Spain); SNSF and SER (Switzerland);
NASU (Ukraine); STFC (United Kingdom); NSF (USA).
We acknowledge the computing resources that are provided by CERN, IN2P3 (France), KIT and DESY (Germany), INFN (Italy), SURF (The Netherlands), PIC (Spain), GridPP (United Kingdom), RRCKI and Yandex LLC (Russia), CSCS (Switzerland), IFIN-HH (Romania), CBPF (Brazil), PL-GRID (Poland) and OSC (USA). We are indebted to the communities behind the multiple open
source software packages on which we depend.
Individual groups or members have received support from AvH Foundation (Germany),
EPLANET, Marie Sk\l{}odowska-Curie Actions and ERC (European Union),
Conseil G\'{e}n\'{e}ral de Haute-Savoie, Labex ENIGMASS and OCEVU,
R\'{e}gion Auvergne (France), RFBR and Yandex LLC (Russia), GVA, XuntaGal and GENCAT (Spain), Herchel Smith Fund, The Royal Society, Royal Commission for the Exhibition of 1851 and the Leverhulme Trust (United Kingdom).

\addcontentsline{toc}{section}{References}
\ifx\mcitethebibliography\mciteundefinedmacro
\PackageError{LHCb.bst}{mciteplus.sty has not been loaded}
{This bibstyle requires the use of the mciteplus package.}\fi
\providecommand{\href}[2]{#2}

\newpage

\centerline{\large\bf LHCb collaboration}
\begin{flushleft}
\small
R.~Aaij$^{39}$,
B.~Adeva$^{38}$,
M.~Adinolfi$^{47}$,
Z.~Ajaltouni$^{5}$,
S.~Akar$^{6}$,
J.~Albrecht$^{10}$,
F.~Alessio$^{39}$,
M.~Alexander$^{52}$,
S.~Ali$^{42}$,
G.~Alkhazov$^{31}$
% ...

{\footnotesize \it
% ...
$ ^{5}$Clermont Universit{\'e}, Universit{\'e} Blaise Pascal, CNRS/IN2P3, LPC, Clermont-Ferrand, France\\
$ ^{6}$CPPM, Aix-Marseille Universit{\'e}, CNRS/IN2P3, Marseille, France\\
% ...
$ ^{10}$Fakult{\"a}t Physik, Technische Universit{\"a}t Dortmund, Dortmund, Germany\\
% ...
$ ^{31}$Petersburg Nuclear Physics Institute (PNPI), Gatchina, Russia\\
% ...
$ ^{38}$Universidad de Santiago de Compostela, Santiago de Compostela, Spain\\
$ ^{39}$European Organization for Nuclear Research (CERN), Geneva, Switzerland\\
% ...
$ ^{42}$Nikhef National Institute for Subatomic Physics, Amsterdam, The Netherlands\\
% ...
$ ^{47}$H.H. Wills Physics Laboratory, University of Bristol, Bristol, United Kingdom\\
% ...
$ ^{52}$School of Physics and Astronomy, University of Glasgow, Glasgow, United Kingdom\\
% ...
}
\end{flushleft}


\begin{mcitethebibliography}{10}
\mciteSetBstSublistMode{n}
\mciteSetBstMaxWidthForm{subitem}{\alph{mcitesubitemcount})}
\mciteSetBstSublistLabelBeginEnd{\mcitemaxwidthsubitemform\space}
{\relax}{\relax}

\bibitem{LHCb-PAPER-2012-031}
LHCb collaboration, {R.\ Aaij \emph{et al.\ }, and A.\ Bharucha} {\em et~al.},
  \ifthenelse{\boolean{articletitles}}{\emph{{Implications of LHCb measurements
  and future prospects}},
  }{}\href{http://dx.doi.org/10.1140/epjc/s10052-013-2373-2}{Eur.\ Phys.\ J.\
  \textbf{C73} (2013) 2373},
  \href{http://arxiv.org/abs/1208.3355}{{\normalfont\ttfamily
  arXiv:1208.3355}}\relax
\mciteBstWouldAddEndPuncttrue
\mciteSetBstMidEndSepPunct{\mcitedefaultmidpunct}
{\mcitedefaultendpunct}{\mcitedefaultseppunct}\relax
\EndOfBibitem
\bibitem{Fleischer:2011cw}
R.~Fleischer and R.~Knegjens,
  \ifthenelse{\boolean{articletitles}}{\emph{{Effective lifetimes of $B_s$
  decays and their constraints on the $B_s^0$-$\overline{B}^0_s$ mixing
  parameters}},
  }{}\href{http://dx.doi.org/10.1140/epjc/s10052-011-1789-9}{Eur.\ Phys.\ J.\
  \textbf{C71} (2011) 1789},
  \href{http://arxiv.org/abs/1109.5115}{{\normalfont\ttfamily
  arXiv:1109.5115}}\relax
\mciteBstWouldAddEndPuncttrue
\mciteSetBstMidEndSepPunct{\mcitedefaultmidpunct}
{\mcitedefaultendpunct}{\mcitedefaultseppunct}\relax
\EndOfBibitem
\bibitem{Fleischer:2011ib}
R.~Fleischer, R.~Knegjens, and G.~Ricciardi,
  \ifthenelse{\boolean{articletitles}}{\emph{{Exploring CP violation and
  $\eta$-$\eta'$ mixing with the $B^0_{s,d} \to J/\psi \eta^{(\prime)}$
  systems}}, }{}\href{http://dx.doi.org/10.1140/epjc/s10052-011-1798-8}{Eur.\
  Phys.\ J.\  \textbf{C71} (2011) 1798},
  \href{http://arxiv.org/abs/1110.5490}{{\normalfont\ttfamily
  arXiv:1110.5490}}\relax
\mciteBstWouldAddEndPuncttrue
\mciteSetBstMidEndSepPunct{\mcitedefaultmidpunct}
{\mcitedefaultendpunct}{\mcitedefaultseppunct}\relax
\EndOfBibitem
\bibitem{LHCb-PAPER-2014-059}
LHCb collaboration, R.~Aaij {\em et~al.},
  \ifthenelse{\boolean{articletitles}}{\emph{{Precision measurement of $C\!P$
  violation in $B_s^0 \to J/\psi K^+K^-$ decays}},
  }{}\href{http://dx.doi.org/10.1103/PhysRevLett.114.041801}{Phys.\ Rev.\
  Lett.\  \textbf{114} (2015) 041801},
  \href{http://arxiv.org/abs/1411.3104}{{\normalfont\ttfamily
  arXiv:1411.3104}}\relax
\mciteBstWouldAddEndPuncttrue
\mciteSetBstMidEndSepPunct{\mcitedefaultmidpunct}
{\mcitedefaultendpunct}{\mcitedefaultseppunct}\relax
\EndOfBibitem
\bibitem{LHCb-PAPER-2014-019}
LHCb collaboration, R.~Aaij {\em et~al.},
  \ifthenelse{\boolean{articletitles}}{\emph{{Measurement of the
  $C\!P$-violating phase $\phi_s$ in $\overline{B}^0_s\to J/\psi\pi^+\pi^-$
  decays}}, }{}\href{http://dx.doi.org/10.1016/j.physletb.2014.06.079}{Phys.\
  Lett.\  \textbf{B736} (2014) 186},
  \href{http://arxiv.org/abs/1405.4140}{{\normalfont\ttfamily
  arXiv:1405.4140}}\relax
\mciteBstWouldAddEndPuncttrue
\mciteSetBstMidEndSepPunct{\mcitedefaultmidpunct}
{\mcitedefaultendpunct}{\mcitedefaultseppunct}\relax
\EndOfBibitem
\bibitem{Lenz:2012mb}
A.~Lenz, \ifthenelse{\boolean{articletitles}}{\emph{{Theoretical update of
  $B$-Mixing and lifetimes}}, }{} in {\em {2012 Electroweak Interactions and
  Unified Theories}}, Moriond, 2012.
\newblock \href{http://arxiv.org/abs/1205.1444}{{\normalfont\ttfamily
  arXiv:1205.1444}}\relax
\mciteBstWouldAddEndPuncttrue
\mciteSetBstMidEndSepPunct{\mcitedefaultmidpunct}
{\mcitedefaultendpunct}{\mcitedefaultseppunct}\relax
\EndOfBibitem
\bibitem{LHCb-PAPER-2013-060}
LHCb collaboration, R.~Aaij {\em et~al.},
  \ifthenelse{\boolean{articletitles}}{\emph{{Measurement of the
  $\overline{B}^0_s\to D_s^-D_s^+$ and $\overline{B}^0_s\to D^-D_s^+$ effective
  lifetimes}},
  }{}\href{http://dx.doi.org/10.1103/PhysRevLett.112.111802}{Phys.\ Rev.\
  Lett.\  \textbf{112} (2014) 111802},
  \href{http://arxiv.org/abs/1312.1217}{{\normalfont\ttfamily
  arXiv:1312.1217}}\relax
\mciteBstWouldAddEndPuncttrue
\mciteSetBstMidEndSepPunct{\mcitedefaultmidpunct}
{\mcitedefaultendpunct}{\mcitedefaultseppunct}\relax
\EndOfBibitem
\bibitem{LHCb-PAPER-2014-011}
LHCb collaboration, R.~Aaij {\em et~al.},
  \ifthenelse{\boolean{articletitles}}{\emph{{Effective lifetime measurements
  in the $B_s^0\to K^+K^-$, $B^0\to K^+\pi^-$ and $B_s^0\to \pi^+K^-$ decays}},
  }{}\href{http://dx.doi.org/10.1016/j.physletb.2014.07.051}{Phys.\ Lett.\
  \textbf{B736} (2014) 446},
  \href{http://arxiv.org/abs/1406.7204}{{\normalfont\ttfamily
  arXiv:1406.7204}}\relax
\mciteBstWouldAddEndPuncttrue
\mciteSetBstMidEndSepPunct{\mcitedefaultmidpunct}
{\mcitedefaultendpunct}{\mcitedefaultseppunct}\relax
\EndOfBibitem
\bibitem{Alves:2008zz}
LHCb collaboration, A.~A. Alves~Jr.\ {\em et~al.},
  \ifthenelse{\boolean{articletitles}}{\emph{{The \lhcb detector at the LHC}},
  }{}\href{http://dx.doi.org/10.1088/1748-0221/3/08/S08005}{JINST \textbf{3}
  (2008) S08005}\relax
\mciteBstWouldAddEndPuncttrue
\mciteSetBstMidEndSepPunct{\mcitedefaultmidpunct}
{\mcitedefaultendpunct}{\mcitedefaultseppunct}\relax
\EndOfBibitem
\bibitem{LHCb-DP-2014-002}
LHCb collaboration, R.~Aaij {\em et~al.},
  \ifthenelse{\boolean{articletitles}}{\emph{{LHCb detector performance}},
  }{}\href{http://dx.doi.org/10.1142/S0217751X15300227}{Int.\ J.\ Mod.\ Phys.\
  \textbf{A30} (2015) 1530022},
  \href{http://arxiv.org/abs/1412.6352}{{\normalfont\ttfamily
  arXiv:1412.6352}}\relax
\mciteBstWouldAddEndPuncttrue
\mciteSetBstMidEndSepPunct{\mcitedefaultmidpunct}
{\mcitedefaultendpunct}{\mcitedefaultseppunct}\relax
\EndOfBibitem
\bibitem{LHCb-PAPER-2012-048}
LHCb collaboration, R.~Aaij {\em et~al.},
  \ifthenelse{\boolean{articletitles}}{\emph{{Measurements of the
  $\Lambda_b^0$, $\Xi_b^-$, and $\Omega_b^-$ baryon masses}},
  }{}\href{http://dx.doi.org/10.1103/PhysRevLett.110.182001}{Phys.\ Rev.\
  Lett.\  \textbf{110} (2013) 182001},
  \href{http://arxiv.org/abs/1302.1072}{{\normalfont\ttfamily
  arXiv:1302.1072}}\relax
\mciteBstWouldAddEndPuncttrue
\mciteSetBstMidEndSepPunct{\mcitedefaultmidpunct}
{\mcitedefaultendpunct}{\mcitedefaultseppunct}\relax
\EndOfBibitem
\bibitem{LHCb-DP-2012-004}
R.~Aaij {\em et~al.}, \ifthenelse{\boolean{articletitles}}{\emph{{The \lhcb
  trigger and its performance in 2011}},
  }{}\href{http://dx.doi.org/10.1088/1748-0221/8/04/P04022}{JINST \textbf{8}
  (2013) P04022}, \href{http://arxiv.org/abs/1211.3055}{{\normalfont\ttfamily
  arXiv:1211.3055}}\relax
\mciteBstWouldAddEndPuncttrue
\mciteSetBstMidEndSepPunct{\mcitedefaultmidpunct}
{\mcitedefaultendpunct}{\mcitedefaultseppunct}\relax
\EndOfBibitem
\bibitem{Sjostrand:2006za}
T.~Sj\"{o}strand, S.~Mrenna, and P.~Skands,
  \ifthenelse{\boolean{articletitles}}{\emph{{PYTHIA 6.4 physics and manual}},
  }{}\href{http://dx.doi.org/10.1088/1126-6708/2006/05/026}{JHEP \textbf{05}
  (2006) 026}, \href{http://arxiv.org/abs/hep-ph/0603175}{{\normalfont\ttfamily
  arXiv:hep-ph/0603175}}\relax
\mciteBstWouldAddEndPuncttrue
\mciteSetBstMidEndSepPunct{\mcitedefaultmidpunct}
{\mcitedefaultendpunct}{\mcitedefaultseppunct}\relax
\EndOfBibitem
\bibitem{Sjostrand:2007gs}
T.~Sj\"{o}strand, S.~Mrenna, and P.~Skands,
  \ifthenelse{\boolean{articletitles}}{\emph{{A brief introduction to PYTHIA
  8.1}}, }{}\href{http://dx.doi.org/10.1016/j.cpc.2008.01.036}{Comput.\ Phys.\
  Commun.\  \textbf{178} (2008) 852},
  \href{http://arxiv.org/abs/0710.3820}{{\normalfont\ttfamily
  arXiv:0710.3820}}\relax
\mciteBstWouldAddEndPuncttrue
\mciteSetBstMidEndSepPunct{\mcitedefaultmidpunct}
{\mcitedefaultendpunct}{\mcitedefaultseppunct}\relax
\EndOfBibitem
\bibitem{LHCb-PROC-2010-056}
I.~Belyaev {\em et~al.}, \ifthenelse{\boolean{articletitles}}{\emph{{Handling
  of the generation of primary events in Gauss, the LHCb simulation
  framework}}, }{}\href{http://dx.doi.org/10.1088/1742-6596/331/3/032047}{{J.\
  Phys.\ Conf.\ Ser.\ } \textbf{331} (2011) 032047}\relax
\mciteBstWouldAddEndPuncttrue
\mciteSetBstMidEndSepPunct{\mcitedefaultmidpunct}
{\mcitedefaultendpunct}{\mcitedefaultseppunct}\relax
\EndOfBibitem
\bibitem{Lange:2001uf}
D.~J. Lange, \ifthenelse{\boolean{articletitles}}{\emph{{The EvtGen particle
  decay simulation package}},
  }{}\href{http://dx.doi.org/10.1016/S0168-9002(01)00089-4}{Nucl.\ Instrum.\
  Meth.\  \textbf{A462} (2001) 152}\relax
\mciteBstWouldAddEndPuncttrue
\mciteSetBstMidEndSepPunct{\mcitedefaultmidpunct}
{\mcitedefaultendpunct}{\mcitedefaultseppunct}\relax
\EndOfBibitem
\bibitem{Golonka:2005pn}
P.~Golonka and Z.~Was, \ifthenelse{\boolean{articletitles}}{\emph{{PHOTOS Monte
  Carlo: A precision tool for QED corrections in $Z$ and $W$ decays}},
  }{}\href{http://dx.doi.org/10.1140/epjc/s2005-02396-4}{Eur.\ Phys.\ J.\
  \textbf{C45} (2006) 97},
  \href{http://arxiv.org/abs/hep-ph/0506026}{{\normalfont\ttfamily
  arXiv:hep-ph/0506026}}\relax
\mciteBstWouldAddEndPuncttrue
\mciteSetBstMidEndSepPunct{\mcitedefaultmidpunct}
{\mcitedefaultendpunct}{\mcitedefaultseppunct}\relax
\EndOfBibitem
\bibitem{Allison:2006ve}
Geant4 collaboration, J.~Allison {\em et~al.},
  \ifthenelse{\boolean{articletitles}}{\emph{{Geant4 developments and
  applications}}, }{}\href{http://dx.doi.org/10.1109/TNS.2006.869826}{IEEE
  Trans.\ Nucl.\ Sci.\  \textbf{53} (2006) 270}\relax
\mciteBstWouldAddEndPuncttrue
\mciteSetBstMidEndSepPunct{\mcitedefaultmidpunct}
{\mcitedefaultendpunct}{\mcitedefaultseppunct}\relax
\EndOfBibitem
\bibitem{Agostinelli:2002hh}
Geant4 collaboration, S.~Agostinelli {\em et~al.},
  \ifthenelse{\boolean{articletitles}}{\emph{{Geant4: A simulation toolkit}},
  }{}\href{http://dx.doi.org/10.1016/S0168-9002(03)01368-8}{Nucl.\ Instrum.\
  Meth.\  \textbf{A506} (2003) 250}\relax
\mciteBstWouldAddEndPuncttrue
\mciteSetBstMidEndSepPunct{\mcitedefaultmidpunct}
{\mcitedefaultendpunct}{\mcitedefaultseppunct}\relax
\EndOfBibitem
\bibitem{LHCb-PROC-2011-006}
M.~Clemencic {\em et~al.}, \ifthenelse{\boolean{articletitles}}{\emph{{The
  \lhcb simulation application, Gauss: Design, evolution and experience}},
  }{}\href{http://dx.doi.org/10.1088/1742-6596/331/3/032023}{{J.\ Phys.\ Conf.\
  Ser.\ } \textbf{331} (2011) 032023}\relax
\mciteBstWouldAddEndPuncttrue
\mciteSetBstMidEndSepPunct{\mcitedefaultmidpunct}
{\mcitedefaultendpunct}{\mcitedefaultseppunct}\relax
\EndOfBibitem
\bibitem{PDG2014}
Particle Data Group, K.~A. Olive {\em et~al.},
  \ifthenelse{\boolean{articletitles}}{\emph{{\href{http://pdg.lbl.gov/}{Review
  of particle physics}}},
  }{}\href{http://dx.doi.org/10.1088/1674-1137/38/9/090001}{Chin.\ Phys.\
  \textbf{C38} (2014) 090001}, {and 2015 update}\relax
\mciteBstWouldAddEndPuncttrue
\mciteSetBstMidEndSepPunct{\mcitedefaultmidpunct}
{\mcitedefaultendpunct}{\mcitedefaultseppunct}\relax
\EndOfBibitem
\bibitem{Hulsbergen:2005pu}
W.~D. Hulsbergen, \ifthenelse{\boolean{articletitles}}{\emph{{Decay chain
  fitting with a Kalman filter}},
  }{}\href{http://dx.doi.org/10.1016/j.nima.2005.06.078}{Nucl.\ Instrum.\
  Meth.\  \textbf{A552} (2005) 566},
  \href{http://arxiv.org/abs/physics/0503191}{{\normalfont\ttfamily
  arXiv:physics/0503191}}\relax
\mciteBstWouldAddEndPuncttrue
\mciteSetBstMidEndSepPunct{\mcitedefaultmidpunct}
{\mcitedefaultendpunct}{\mcitedefaultseppunct}\relax
\EndOfBibitem
\bibitem{LHCb-PAPER-2013-065}
LHCb collaboration, R.~Aaij {\em et~al.},
  \ifthenelse{\boolean{articletitles}}{\emph{{Measurements of the $B^+$, $B^0$,
  $B_s^0$ meson and $\Lambda_b^0$ baryon lifetimes}},
  }{}\href{http://dx.doi.org/10.1007/JHEP04(2014)114}{JHEP \textbf{04} (2014)
  114}, \href{http://arxiv.org/abs/1402.2554}{{\normalfont\ttfamily
  arXiv:1402.2554}}\relax
\mciteBstWouldAddEndPuncttrue
\mciteSetBstMidEndSepPunct{\mcitedefaultmidpunct}
{\mcitedefaultendpunct}{\mcitedefaultseppunct}\relax
\EndOfBibitem
\bibitem{Hocker:2007ht}
A.~Hoecker {\em et~al.}, \ifthenelse{\boolean{articletitles}}{\emph{{TMVA:
  Toolkit for multivariate data analysis}}, }{}PoS \textbf{ACAT} (2007) 040,
  \href{http://arxiv.org/abs/physics/0703039}{{\normalfont\ttfamily
  arXiv:physics/0703039}}\relax
\mciteBstWouldAddEndPuncttrue
\mciteSetBstMidEndSepPunct{\mcitedefaultmidpunct}
{\mcitedefaultendpunct}{\mcitedefaultseppunct}\relax
\EndOfBibitem
\bibitem{Bukin}
BABAR collaboration, J.~P. Lees {\em et~al.},
  \ifthenelse{\boolean{articletitles}}{\emph{{Branching fraction measurements
  of the color-suppressed decays $\overline{B}^0 \to D^{(*)0} \pi^0$, $D^{(*)0}
  \eta$, $D^{(*)0} \omega$, and $D^{(*)0} \eta^\prime$ and measurement of the
  polarization in the decay $\overline{B}^0 \to D^{*0} \omega$}},
  }{}\href{http://dx.doi.org/10.1103/PhysRevD.84.112007,
  10.1103/PhysRevD.87.039901}{Phys.\ Rev.\  \textbf{D84} (2011) 112007},
  \href{http://arxiv.org/abs/1107.5751}{{\normalfont\ttfamily
  arXiv:1107.5751}}, [Erratum: Phys. Rev.D87,039901(2013)]\relax
\mciteBstWouldAddEndPuncttrue
\mciteSetBstMidEndSepPunct{\mcitedefaultmidpunct}
{\mcitedefaultendpunct}{\mcitedefaultseppunct}\relax
\EndOfBibitem
\bibitem{JOHNSON01061949}
N.~L. Johnson, \ifthenelse{\boolean{articletitles}}{\emph{Systems of frequency
  curves generated by methods of translation},
  }{}\href{http://dx.doi.org/10.1093/biomet/36.1-2.149}{Biometrika \textbf{36}
  (1949), no.~1-2 149}\relax
\mciteBstWouldAddEndPuncttrue
\mciteSetBstMidEndSepPunct{\mcitedefaultmidpunct}
{\mcitedefaultendpunct}{\mcitedefaultseppunct}\relax
\EndOfBibitem
\bibitem{LHCb-PAPER-2015-010}
LHCb collaboration, R.~Aaij {\em et~al.},
  \ifthenelse{\boolean{articletitles}}{\emph{{Observation of the decay
  $\overline{B}^0_s\to\psi(2S)K^+\pi^-$}},
  }{}\href{http://dx.doi.org/10.1016/j.physletb.2015.06.038}{Phys.\ Lett.\
  \textbf{B747} (2015) 484},
  \href{http://arxiv.org/abs/1503.07112}{{\normalfont\ttfamily
  arXiv:1503.07112}}\relax
\mciteBstWouldAddEndPuncttrue
\mciteSetBstMidEndSepPunct{\mcitedefaultmidpunct}
{\mcitedefaultendpunct}{\mcitedefaultseppunct}\relax
\EndOfBibitem
\bibitem{Chang:2006sd}
Belle collaboration, M.~C. Chang {\em et~al.},
  \ifthenelse{\boolean{articletitles}}{\emph{{Observation of the decay $B^0
  \rightarrow J/\psi \eta$}},
  }{}\href{http://dx.doi.org/10.1103/PhysRevLett.98.131803}{Phys.\ Rev.\ Lett.\
   \textbf{98} (2007) 131803},
  \href{http://arxiv.org/abs/hep-ex/0609047}{{\normalfont\ttfamily
  arXiv:hep-ex/0609047}}\relax
\mciteBstWouldAddEndPuncttrue
\mciteSetBstMidEndSepPunct{\mcitedefaultmidpunct}
{\mcitedefaultendpunct}{\mcitedefaultseppunct}\relax
\EndOfBibitem
\bibitem{Chang:2012gnb}
Belle collaboration, M.~C. Chang {\em et~al.},
  \ifthenelse{\boolean{articletitles}}{\emph{{Measurement of $B^0 \to J/\psi
  \eta^{(\prime)}$ and constraint on the $\eta-\eta^\prime$ mixing angle}},
  }{}\href{http://dx.doi.org/10.1103/PhysRevD.85.091102}{Phys.\ Rev.\
  \textbf{D85} (2012) 091102},
  \href{http://arxiv.org/abs/1203.3399}{{\normalfont\ttfamily
  arXiv:1203.3399}}\relax
\mciteBstWouldAddEndPuncttrue
\mciteSetBstMidEndSepPunct{\mcitedefaultmidpunct}
{\mcitedefaultendpunct}{\mcitedefaultseppunct}\relax
\EndOfBibitem
\bibitem{LHCb-PAPER-2014-056}
LHCb collaboration, R.~Aaij {\em et~al.},
  \ifthenelse{\boolean{articletitles}}{\emph{{Study of $\eta$--$\eta^{\prime}$
  mixing from measurement of $B^0_{(s)} \to J/\psi \eta^{(\prime)}$ decay
  rates}}, }{}\href{http://dx.doi.org/10.1007/JHEP01(2015)024}{JHEP \textbf{01}
  (2015) 024}, \href{http://arxiv.org/abs/1411.0943}{{\normalfont\ttfamily
  arXiv:1411.0943}}\relax
\mciteBstWouldAddEndPuncttrue
\mciteSetBstMidEndSepPunct{\mcitedefaultmidpunct}
{\mcitedefaultendpunct}{\mcitedefaultseppunct}\relax
\EndOfBibitem
\bibitem{LHCb-PAPER-2011-018}
LHCb collaboration, R.~Aaij {\em et~al.},
  \ifthenelse{\boolean{articletitles}}{\emph{{Measurement of $b$ hadron
  production fractions in 7 TeV $pp$ collisions}},
  }{}\href{http://dx.doi.org/10.1103/PhysRevD.85.032008}{Phys.\ Rev.\
  \textbf{D85} (2012) 032008},
  \href{http://arxiv.org/abs/1111.2357}{{\normalfont\ttfamily
  arXiv:1111.2357}}\relax
\mciteBstWouldAddEndPuncttrue
\mciteSetBstMidEndSepPunct{\mcitedefaultmidpunct}
{\mcitedefaultendpunct}{\mcitedefaultseppunct}\relax
\EndOfBibitem
\bibitem{LHCb-PAPER-2012-037}
LHCb collaboration, R.~Aaij {\em et~al.},
  \ifthenelse{\boolean{articletitles}}{\emph{{Measurement of the fragmentation
  fraction ratio $f_s/f_d$ and its dependence on $B$ meson kinematics}},
  }{}\href{http://dx.doi.org/10.1007/JHEP04(2013)001}{JHEP \textbf{04} (2013)
  001}, \href{http://arxiv.org/abs/1301.5286}{{\normalfont\ttfamily
  arXiv:1301.5286}}\relax
\mciteBstWouldAddEndPuncttrue
\mciteSetBstMidEndSepPunct{\mcitedefaultmidpunct}
{\mcitedefaultendpunct}{\mcitedefaultseppunct}\relax
\EndOfBibitem
\bibitem{LHCb-CONF-2013-011}
{LHCb collaboration}, \ifthenelse{\boolean{articletitles}}{\emph{{Updated
  average $f_{s}/f_{d}$ $b$-hadron production fraction ratio for $7 \tev$ $pp$
  collisions}}, }{}
  \href{http://cdsweb.cern.ch/search?p=LHCb-CONF-2013-011&f=reportnumber&action_search=Search&c=LHCb+Conference+Contributions}
  {LHCb-CONF-2013-011}\relax
\mciteBstWouldAddEndPuncttrue
\mciteSetBstMidEndSepPunct{\mcitedefaultmidpunct}
{\mcitedefaultendpunct}{\mcitedefaultseppunct}\relax
\EndOfBibitem
\bibitem{Ikeda:1999aq}
Belle collaboration, H.~Ikeda {\em et~al.},
  \ifthenelse{\boolean{articletitles}}{\emph{{A detailed test of the CsI(Tl)
  calorimeter for BELLE with photon beams of energy between 20-MeV and
  5.4-GeV}}, }{}\href{http://dx.doi.org/10.1016/S0168-9002(99)00992-4}{Nucl.\
  Instrum.\ Meth.\  \textbf{A441} (2000) 401}\relax
\mciteBstWouldAddEndPuncttrue
\mciteSetBstMidEndSepPunct{\mcitedefaultmidpunct}
{\mcitedefaultendpunct}{\mcitedefaultseppunct}\relax
\EndOfBibitem
\bibitem{Dauncey:2014xga}
P.~D. Dauncey {\em et~al.},
  \ifthenelse{\boolean{articletitles}}{\emph{{Handling uncertainties in
  background shapes: the discrete profiling method}},
  }{}\href{http://dx.doi.org/10.1088/1748-0221/10/04/P04015}{JINST \textbf{10}
  (2015) P04015}, \href{http://arxiv.org/abs/1408.6865}{{\normalfont\ttfamily
  arXiv:1408.6865}}\relax
\mciteBstWouldAddEndPuncttrue
\mciteSetBstMidEndSepPunct{\mcitedefaultmidpunct}
{\mcitedefaultendpunct}{\mcitedefaultseppunct}\relax
\EndOfBibitem
\bibitem{HFAG}
Heavy Flavor Averaging Group, Y.~Amhis {\em et~al.},
  \ifthenelse{\boolean{articletitles}}{\emph{{Averages of $b$-hadron,
  $c$-hadron, and $\tau$-lepton properties as of summer 2014}},
  }{}\href{http://arxiv.org/abs/1412.7515}{{\normalfont\ttfamily
  arXiv:1412.7515}}, {updated results and plots available at
  \href{http://www.slac.stanford.edu/xorg/hfag/}{{\texttt{http://www.slac.stanford.edu/xorg/hfag/}}}}\relax
\mciteBstWouldAddEndPuncttrue
\mciteSetBstMidEndSepPunct{\mcitedefaultmidpunct}
{\mcitedefaultendpunct}{\mcitedefaultseppunct}\relax
\EndOfBibitem
\end{mcitethebibliography}
\end{document}